\documentclass[12pt]{article}
\usepackage{epsfig,amsfonts,amssymb}
\usepackage{cite}
\input epsf.sty
\usepackage{cite}

\def\ZZZ{{\hbox{ Z\kern-1.6mm Z}}}
\def\RRR{{\hbox{ R\kern-2.4mm R}}}
\def\CCC{{\hbox{ C\kern-2.0mm C}}}
\def\zzz{{\hbox{z\kern-1mm z}}}

\newcommand{\qeq}{{\hbox{=\kern-2.3mm ? \kern.5mm }}}
\renewcommand{\qeq}{=}

\newcommand{\ws}{{\wt\sigma}}
\newcommand{\wrh}{{\wt\rho}}
\newcommand{\wv}{{\wt v}}

\newcommand{\LL}{{\cal L}}

\newcommand{\wt}{\widetilde}
\newcommand{\wh}{\widehat}
\newcommand{\wc}{\wt}
\newcommand{\wb}{\bar}

\newcommand{\RR}{{\cal R}}
\newcommand{\NN}{{\cal N}}

\newcommand{\be}{\begin{equation}}
\newcommand{\ee}{\end{equation}}
\newcommand{\ben}{\begin{eqnarray}\displaystyle}
\newcommand{\een}{\end{eqnarray}}

\newcommand{\bea}[1]{\begin{eqnarray}\label{#1} }
\newcommand{\eea}{\end{eqnarray}}

\newcommand{\refb}[1]{(\ref{#1})}

\renewcommand{\vec}{}

\def\one{{\hbox{ 1\kern-.8mm l}}}
\def\zero{{\hbox{ 0\kern-1.5mm 0}}}

\def\wa{{\wh a}}
\def\wb{{\wh b}}
\def\wc{{\wh c}}
\def\wdd{{\wh d}}

\begin{document}

\begin{center}
{\Large \bf
Three String Junction and $\NN=4$ Dyon Spectrum}

\end{center}

\vskip .6cm
\medskip

\vspace*{4.0ex}

\centerline{\large \rm  }

\vspace*{4.0ex}

\centerline{\large \rm Ashoke Sen}

\vspace*{4.0ex}

\centerline{\large \it Harish-Chandra Research Institute}

\centerline{\large \it  Chhatnag Road, Jhusi,
Allahabad 211019, INDIA}

\vspace*{1.0ex}
\centerline{E-mail:  sen@mri.ernet.in}

\vspace*{5.0ex}

\centerline{\bf Abstract} \bigskip

The exact spectrum of dyons in a class of $\NN=4$ supersymmetric 
string theories gives us information about dyon spectrum in
$\NN=4$ supersymmetric gauge theories. This in turn can be translated
into prediction about the BPS spectrum of three string junctions
on a configuration of three parallel D3-branes. 
We show that
this prediction agrees with the known spectrum of three string
junction in different domains
in the moduli space separated by walls of marginal stability.

\vfill \eject

\baselineskip=18pt



We now have a good understanding
of the exact spectrum of a class of
quarter BPS dyons in a variety of $\NN=4$
supersymmetric string 
theories\cite{9607026,0412287,0505094,0506249,0508174,
0510147,0602254,
0603066,0605210,0607155,0609109,0612011,0702141,
0702150,0705.1433,0705.3874,0706.2363,0708.1270}. 
Since by going to the appropriate
region in the moduli space of these string theories
and taking a decoupling limit we can
recover $\NN=4$ supersymmetric gauge 
theories\cite{narain,nsw}, the
spectrum of quarter BPS dyons in $\NN=4$ supersymmetric
string theories provides us information about the spectrum of
quarter BPS dyons in $\NN=4$ supersymmetric gauge 
theories.
The latter, in turn, can be related to the BPS spectrum of string
junctions 
on  a set of parallel D3 branes\cite{9712211}. 
Thus the known
dyon spectrum in $\NN=4$ supersymmetric string theories gives
us prediction about the BPS spectrum of string junctions
on a set of parallel D3-branes. Our goal is to verify if this
prediction is consistent with the known properties
of string junctions.

We shall work with 
heterotic string theory on 
$T^4\times T^2$ and focus on the four U(1) gauge fields
associated with the components of the metric and 2-form field
along the $T^2$ directions.   The electric
charges associated with these gauge fields are the momenta
$\wh n$ and $n'$ and the fundamental string winding charges
$-\wh w$ and $-w'$ along the two circles of $T^2$, and the magnetic
charges associated with these gauge fields are the H-monopole
charges $-\wh W$ and $-W'$ and the
Kaluza-Klein monopole charges $\wh N$ and $N'$ along the same
two circles. Following the notations and
conventions of \cite{0708.1270}
we define the electric and magnetic charge vectors
as:
\be\label{e2dcharge}
Q=\pmatrix{\wh n\cr n'\cr \wh w\cr w'}, \qquad 
P = \pmatrix{\wh W\cr W'\cr
\wh N\cr N'}\, .
\ee
The complex structure  
and the (complexified) Kahler moduli
of  the torus $T^2$ are
encoded in a $4\times 4$ matrix $M$ satisfying
\be \label{edefm}
M^T L M = L, \quad M^T = M\, ,
\ee
where
\be \label{edefl}
L = \pmatrix{0 & I_2\cr I_2 & 0}\, .
\ee
$I_k$ is the $k\times k$ identity matrix.
The other complex
modulus relevant for our discussion is the
axion-dilaton modulus $\tau=a+iS$, where $a$ is the field obtained
by dualizing the 2-form field in four dimensions and
$S=e^{-2\phi}$, $\phi$ being the dilaton field.

The T-duality transformations associated with $T^2$ 
are generated by $4\times 4$ matrices $\Omega$ with
integer entries and satisfying $\Omega L \Omega^T=L$.
They act on the charges and the
moduli as
\be \label{etdual}
Q\to (\Omega^T)^{-1} Q, \quad P \to (\Omega^T)^{-1} P,
\quad M \to \Omega M \Omega^T, \quad \tau \to \tau\, .
\ee
Thus the combinations
\be \label{ecomb}
Q^2 = Q^T L Q, \quad P^2 = P^T L P, \quad Q\cdot P
= Q^T L P\, 
\ee
are  T-duality invariant.
On the other hand the S-duality transformations are generated
by $SL(2,\ZZZ)$ matrices $\pmatrix{\wa & \wb
\cr \wc & \wdd}$ with
$\wa,\wb,\wc,\wdd
\in\ZZZ$, $\wa\wdd-\wb\wc=1$, 
and act on the charges and the
moduli as
\be \label{esdual}
Q\to \wa Q + \wb P, \quad P\to \wc Q + 
\wdd P\,, \quad
\tau \to {\wa \tau + \wb\over 
\wc\tau + \wdd}, \quad
M\to M\, .
\ee

As was reviewed in \cite{0708.1270}, for all charge vectors
$(Q,P)$ which are related to the charge vectors
\be \label{eseed}
Q = \pmatrix{k_3\cr k_4 \cr k_5 \cr -1}, \qquad
P = \pmatrix{l_3\cr l_4 \cr l_5 \cr 0}, \qquad  
\quad k_i, l_i\in \ZZZ 
\, , \qquad \hbox{g.c.d.}(l_3,l_5)=1\, ,
\ee 
by a T-duality transformation, we have a simple formula for the
degeneracy  (more precisely the number of bosonic supermultiplets
minus the number of fermionic supermultiplets) of quarter BPS states:
\be\label{edeg1}
d(\vec Q,\vec P) = (-1)^{Q\cdot P+1}\, g\left({1\over 2}Q^2, 
{1\over 2} P^2, Q\cdot P\right)\, ,
\ee
where $g(m,n,p)$ are defined as the coefficients of Fourier expansion
of a known function $1/\wt\Phi$:
\be \label{edeg2}
{1
\over \wt\Phi(\wt \rho,\wt \sigma, \wt v)}
= \sum_{m,n,p\in \zzz\atop m\ge -1, n\ge -1} g(m,n,p) \,
e^{2\pi i (m\wt\rho+n\wt\sigma+p\wt v)}\, .
\ee
In the particular theory under consideration, $\wt\Phi$ is the well
known Igusa cusp form of weight 
10\cite{igusa1, igusa2, borcherds, 9504006} 
on the moduli space
of genus two Riemann surfaces, parametrized by
the period matrix 
$\pmatrix{\wrh & \wv\cr \wv & \ws}$\cite{9607026}. 
{}From \refb{edeg1}, \refb{edeg2} we see that $d(\vec Q, \vec P)=0$
unless $Q^2\ge -2$ and $P^2\ge -2$.

\begin{figure}
\leavevmode
\begin{center}
\epsfysize=5cm 
\epsfbox{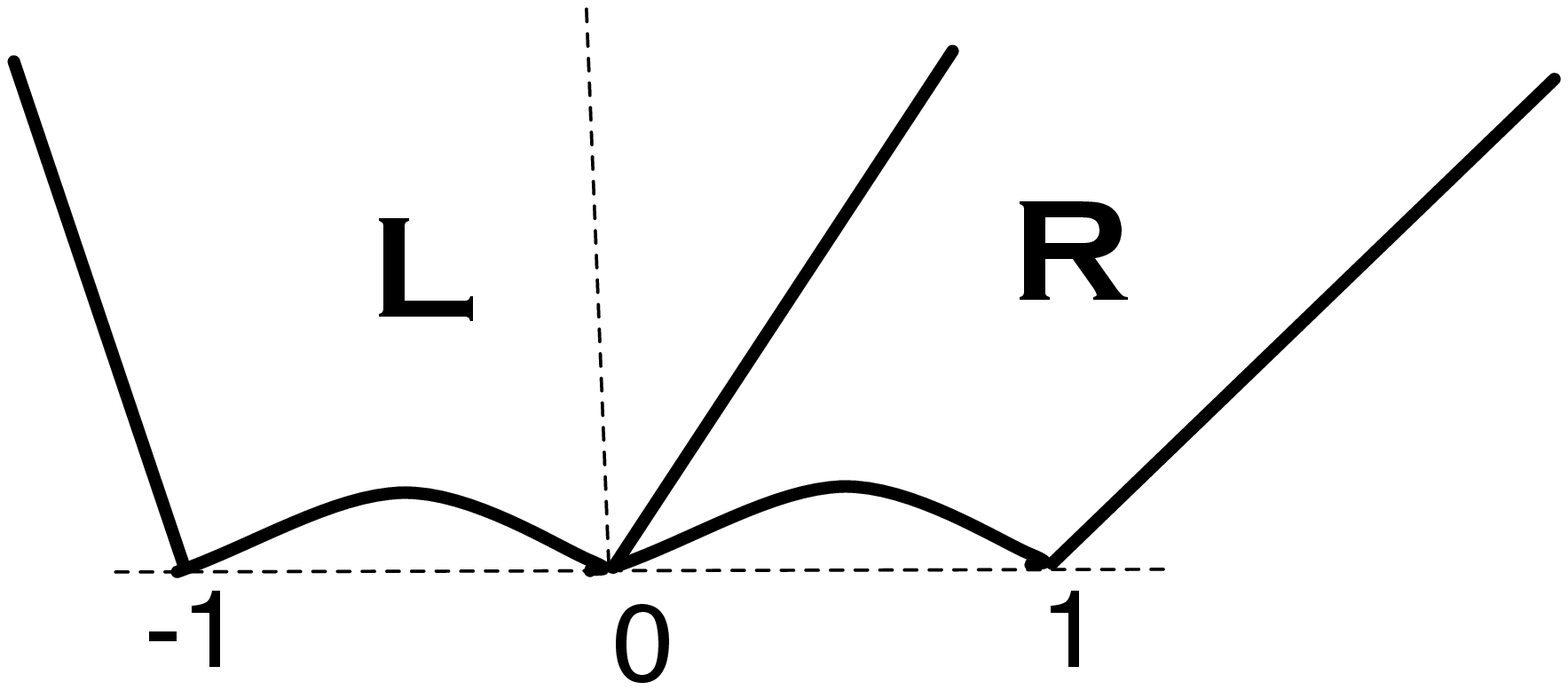}
\end{center}
\vskip -48pt
\caption{The domains $\RR$ and $\LL$.}
\label{f0}
\end{figure}

The formula for the degeneracy given above is not complete unless
we specify the region of the moduli space in which the formula
is valid. As we vary the asymptotic moduli the degeneracy can
actually jump across walls of
marginal stability, -- 
codimension one subspaces of the moduli
space on which the original quarter BPS dyon can decay into a
pair of half-BPS dyons\cite{0702141,0702150,0705.3874,0706.2363,
0707.1563,0707.3035}. 
As has been reviewed 
in detail in
\cite{0708.1270}, a very useful way to label a given wall of
marginal stability is to specify the relation between the
charges of the decay products and the charges of the original
state. In particular the possible decays
of a quarter BPS state with charge $(Q,P)$ are
into half BPS states carrying charges 
$(ad Q-ab P, cd Q - cb P)$ and $(-bc  Q + ab  P, -cd  Q+ad  P)$
with $a,b,c,d\in \ZZZ$, $ad-bc=1$. For fixed values of the
moduli $M$, the corresponding wall is either
a circle in the $\tau$
plane intersecting the real axis at $a/c$ and $b/d$, or
-- for $c=0$ or $d=0$
--  a straight line passing through $b/d$ or $a/c$. The radii of the
circles and the slopes of the straight lines depend on $\vec Q$, $\vec P$
and the other moduli $M$.
The degeneracy formula given in \refb{edeg1}, \refb{edeg2}
is valid inside two separate domains bounded by walls of marginal
stability. The first domain, called $\RR$, is bounded by three
different walls, -- a
straight line through 0,
a circle connecting 0 and 1 and  
a straight line through 1 (see fig.~\ref{f0}). {}From the
decay rules given above it is clear that these three domain walls
correspond to the decays:
\be \label{edecay}
(Q,P) \to (Q, 0) + (0, P), \qquad (Q,P) \to (0, -Q+P) + (Q, Q), \qquad
(Q,P) \to (P, P) + (Q-P, 0) \, .
\ee
The other domain $\LL$ inside which the degeneracy formula is
valid is bounded by a straight line through 0, 
a circle through $-1$ to 0 and a straight line through
$-1$. By following the same rules we see that these
walls correspond to the possible decays:
\be \label{edecay2}
(Q,P) \to (Q, 0) + (0, P), \qquad (Q,P) \to (0, Q+P) + (Q, -Q), \qquad
(Q,P) \to (-P, P) + (Q+P, 0) \, .
\ee

Even though the degeneracy formula 
\refb{edeg1} hold in both domains $\RR$ and
$\LL$, there is a subtle difference between the ways we
extract $g(m,n,p)$ from \refb{edeg2} in the two cases. When we are
computing the formula in the domain $\RR$, we need to expand 
$1/\wt\Phi$ in such a way that for a fixed $m,n$, the sum over $p$
is bounded from above. On the other hand inside the domain $\LL$
we have to expand $1/\wt\Phi$ so that for fixed $m,n$ the sum over
$p$ is bounded from below. 

We shall first focus on the degeneracy
of states with $Q^2=P^2=-2$ and later consider states related to these
by S-duality transformation. 
For this we only need to examine terms
in $1/\wt\Phi$ whose $\ws$, $\wrh$ dependence is of the form
$ e^{-2\pi i \wrh - 2\pi i \ws}$.
The relevant part of $1/\wt\Phi$ is
\be \label{erele}
{1\over \wt\Phi} \simeq e^{-2\pi i \wrh - 2\pi i \ws} {e^{-2\pi i \wv}
\over \left(1 - e^{-2\pi i \wv}\right)^2}\, .
\ee
According to the rule given above we need to expand this in
power of $e^{-2\pi i \wv}$ for calculating the degeneracy in the
domain $\RR$. This gives, in the domain $\RR$,
\be \label{epar1}
d(\vec Q, \vec P) =\cases{0 \quad \hbox{for $Q^2=P^2=-2$,
$Q\cdot P\ge 0$} \cr 
j(-1)^{j-1}
\quad \hbox{for $Q^2=P^2=-2$, $Q\cdot P = -j$, $j>0$}}\, .
\ee
On the other hand in the domain $\LL$ we have to expand 
$1/\wt\Phi$ in powers of $e^{2\pi i \wt v}$ by expressing the
$\wt v$ dependent factor as 
$e^{2\pi i \wt v} / (1 - e^{2\pi i\wt v})^2$. This gives, in 
the domain $\LL$,
\be \label{epar2}
d(\vec Q, \vec P) =\cases{ 0 \quad \hbox{for $Q^2=P^2=-2$,
$Q\cdot P\le 0$}\cr 
j(-1)^{j+1}
\quad \hbox{for $Q^2=P^2=-2$, $Q\cdot P = j$, $j>0$}}\, .
\ee

Let us now focus on a particular state carrying charge vectors
\be \label{egen3}
Q_0 =  \pmatrix{0 \cr 1 \cr 0\cr -1}, 
\qquad P_0  = \pmatrix{-1 \cr 1\cr 1\cr 0}\, .
\ee
This is of the form given in \refb{eseed}, and has
\be \label{egen4}
Q_0^2 = -2, \qquad P_0^2 = -2, \qquad Q_0\cdot P_0 = -1\, .
\ee
Thus according to \refb{epar1}, \refb{epar2}, 
this state will have degeneracy 1 in the
domain $\RR$ and vanishing degeneracy in the domain $\LL$.
In other words, the state will cease to exist as we move from the domain
$\RR$ to the domain $\LL$ crossing the wall separating the two domains.
In the $\tau$ plane this wall is a straight line through 0.

We shall now examine the fate of the state in various other domains.
This is done by noting that the degeneracies in the other domains
may be calculated by mapping them to the domain $\RR$ using an
S-duality transformation and then applying the degeneracy formula
\refb{edeg1}, \refb{edeg2} in the domain $\RR$\cite{0702141}.
Let us consider a domain $\wt\RR$
that is mapped to the domain $\RR$ via
an S-duality transformation matrix $\pmatrix{\wa & \wb \cr \wc & \wdd}$.
This will map the charge vector $(\vec Q_0, \vec P_0)$ to
$(\vec Q_0', \vec P_0')$ given by
\be \label{enewvec}
\vec Q_0' = \wa \vec Q_0 + \wb \vec P_0 = \pmatrix{-\wb \cr \wa+\wb
\cr \wb \cr -\wa}, \qquad 
\vec P_0' = \wc \vec Q_0 + \wdd \vec P_0 
=\pmatrix{-\wdd \cr \wc+\wdd \cr \wdd \cr - \wc}\, .
\ee
Thus $d(\vec Q_0, \vec P_0)$ in the domain $\wt\RR$ is equal
to $d(\vec Q_0',\vec P_0')$ in the domain $\RR$. Although
the charge vectors $(\vec Q_0',\vec P_0')$
do not have the form given in \refb{eseed},
they can be expressed as
\be \label{eobt}
\vec Q_0' = (\Omega^T)^{-1}\vec Q_0'', \qquad
\vec P_0' = (\Omega^T)^{-1} \vec P_0''\, ,
\ee
where
\be \label{eobt2}
Q_0'' = \pmatrix{0\cr \wa^2 +\wa\wb +\wb^2\cr
\wa\wc +\wb \wc +\wb\wdd\cr -1}, \qquad
P_0'' = \pmatrix{-1\cr \wa\wc+\wa\wdd+\wb\wdd\cr \wc^2
+\wc\wdd+\wdd^2\cr 0}, \qquad 
\Omega^T = \pmatrix{\wa & 0 &0&-\wb\cr 0&\wa&\wb&0\cr
0&\wc&\wdd&0\cr -\wc&0&0&\wdd}\, .
\ee
Since $(\vec Q_0''$, $\vec P_0'')$ have the form \refb{eseed}, and
$\Omega$ denotes a T-duality transformation, we conclude that
our degeneracy formula \refb{edeg1}, \refb{edeg2} holds for the
charge vectors \refb{enewvec}. Thus in order to get non-vanishing
$d(\vec Q_0', \vec P_0')$ we must have 
$(Q_0')^2\ge -2$, $(P_0')^2\ge -2$.
Using \refb{enewvec} these conditions translate to
\be \label{etrans}
\wa^2 +\wb^2 +\wa\wb\le 1, \qquad \wc^2 +\wdd^2+\wc\wdd\le 1
\, .
\ee
Since the left hand sides of both equations are positive definite
for $\wa\wdd -\wb\wc=1$, the above equations give strong
constraint on $\wa$, $\wb$, $\wc$ and $\wdd$. In particular for
integer $\wa$, $\wb$, $\wc$, $\wdd$ both
bounds must be saturated. Thus we have $(Q_0')^2=(P_0')^2=-2$.
Eq.\refb{epar1} now tells us that unless $Q_0'\cdot P_0'\le -1$
the degeneracy vanishes in the domain $\RR$. 
This gives rise to one more inequality
\be \label{enewineq}
2(\wa\wc+\wb\wdd) + \wa\wdd+\wb\wc \ge 1\, .
\ee
We can find all integer
solutions to \refb{etrans}, \refb{enewineq} subject to the 
restriction $\wa\wdd-\wb\wc=1$. Up to an overall sign that
does not affect the mapping between the domains in the 
$\tau$ plane, we get the following solutions for
$\pmatrix{\wa &\wb\cr \wc & \wdd}$:
\be \label{elist}
\pmatrix{1 & -1\cr 1 & 0}, \qquad \pmatrix{1 & 0\cr 0 & 1}, 
\qquad \pmatrix{0 & 1\cr -1 & 1}\, .
\ee
This gives the set of all $\wa$, $\wb$, $\wc$ and $\wdd$ for which
$d(\vec Q_0', \vec P_0')$
is non-zero inside $\RR$ and hence $d(\vec Q_0,\vec P_0)$ is
non-zero inside $\wt\RR$. Thus the set of all
domains in which $d(\vec Q_0, \vec P_0)$ is non-zero is obtained
by the image of $\RR$ (for charge vector $(\vec Q_0',\vec P_0')$)
under an S-duality transformation by the
inverse of $\pmatrix{\wa &\wb\cr \wc & \wdd}$.
Now one can easily verify that each of the S-duality transformations
given in \refb{elist}
maps the domain $\RR$ to itself,\footnote{More precisely,
it maps the domain $\RR$ for the
charge vector $(\vec Q_0,\vec P_0)$
to the domain $\RR$ for the charge vector $(Q_0',P_0')$.}  --
this is best seen by noting that
each of these transformations permutes the vertices 0, 1 and $\infty$
of $\RR$. Thus any other domain $\wt\RR$
is mapped to $\RR$ via an
S-duality transformation outside the set \refb{elist},
and hence $d(\vec Q_0,
\vec P_0)$ must vanish in the domain $\wt\RR$.
This leads to the conclusion that
$d(\vec Q_0, \vec P_0)$  vanishes in
all domains outside $\RR$.

We shall now try to verify this prediction by working near a point
in the moduli space where there is an enhanced
$SU(3)$ gauge symmetry. 
This is achieved by taking the matrix valued scalar field $M$ to be
of the form
\be \label{emform}
M = \pmatrix{{4/ 3} & {2/ 3} & -{1/ 3} 
& {2/ 3}\cr
{2/ 3} & {4/ 3} & -{2/ 3} & {1/ 3}\cr
-{1/ 3} & -{2/ 3} & {4/ 3} & -{2/ 3}\cr
{2/ 3} & {1/ 3} & -{2/ 3} & {4/ 3}}\, .
\ee
In this case one finds that
the BPS mass of a purely electric state
vanishes for the
charge vectors
\be \label{eglim}
\alpha = \pmatrix{1 \cr -1 \cr -1\cr 0}, \quad 
\beta = \pmatrix{0 \cr 1\cr 0\cr -1}, \quad
\gamma = \pmatrix{1 \cr 0\cr -1\cr -1}\, .
\ee
Indeed the 
vectors $\pm\alpha$, $\pm\beta$ and $\pm\gamma
=\pm(\alpha+\beta)$ are eigenvectors of $(M+L)$ with zero
eigenvalue and satisfy $\alpha^2=\beta^2=\gamma^2=-1$.
As a result electrically charged states with these charge vectors 
give the six massless electrically charged states which are
necessary for getting the full set of $SU(3)$ gauge fields.
If we adjust the  moduli $M$ to be slightly away from
the one given in \refb{emform} 
then the $SU(3)$ gauge symmetry is spontaneously
broken to $U(1)\times U(1)$
and the charged gauge fields become massive, but remain
light compared to the string scale. 

Besides the
half-BPS massive gauge fields, and the half-BPS dyons related
to these by S-duality transformation, the spontaneously broken
$SU(3)$ gauge theory also
contains quarter BPS dyons carrying electric and magnetic charges
of the form\cite{9712211}
\be \label{equ1}
Q = p \alpha + q\beta, \quad P = r \alpha + s\beta, \quad
p,q,r,s\in \ZZZ\, ,
\ee
in specific domains in the moduli space depending on the
values of $p,q,r,s$.
These must represent some states in the spectrum of quarter
BPS dyons in string theory.\footnote{Note that the gauge theory
limit can be taken by adjusting the moduli $M$ and is
insensitive to $\tau=a+iS$. Thus even after taking the gauge theory
limit we can explore all the different domains separated by walls
of marginal stability by varying $\tau$.}
Conversely every BPS state in string theory, carrying
charge vectors of the form
$(p \alpha + q\beta, r \alpha + s\beta)$,
becomes light compared to the string scale in the region of the moduli
space we are considering, and hence they
must have a realization in gauge theory.
In particular since $(Q_0,P_0)$
defined in \refb{egen3} has the form
\be \label{exx1}
Q_0 = \beta, \qquad P_0 = -\alpha\, ,
\ee
it must have a realization in $SU(3)$ gauge theory.

A simple realization of quarter BPS
dyons in SU(3) gauge theory 
is as a 3-string junction\cite{9607201,9711094} on a system
of three closeby parallel D3-branes\cite{9712211}. 
This system 
has $U(1)\times$
spontaneously broken $SU(3)$ gauge theory as its low 
energy limit.
An $(m,n)$
string ending on a D3-brane carries an electric charge $m$ and
magnetic charge $n$ under the U(1) gauge field living on the
D3-brane. Thus if we have a configuration 
where an $(m_1, n_1)$ string
ends on the first D3-brane, an $(m_2,n_2)$ string ends on the
second D3-brane and an $(m_3, n_3)$ string ends on the 
third D3-brane, with the three strings meeting at a 3-point
junction, then the system will be said to carry 
electric and magnetic charge vectors
\be \label{eem1}
\wt Q = \pmatrix{m_1\cr m_2\cr m_3}\, , \qquad 
\wt P = \pmatrix{n_1\cr n_2\cr n_3}\, .
\ee
Charge conservation at the three string junction
requires $\sum_i m_i$ and $\sum_i n_i$
to vanish. Thus although the gauge theory on the D3-brane  is
$U(3)$, the state described above carries only an $SU(3)$
charge. This allows us to compare a BPS state
of   the configuration described above with that of the $SU(3)$
gauge theory that arises as the low energy limit of heteroric string
theory on $T^4\times T^2$. For this we first need to learn how to
translate a charge vector of the type given in \refb{eem1} to the
one given in \refb{e2dcharge}.  We do this by comparing the
charges carried by the massive gauge fields. On the configuration
of three D3-branes, the massive gauge fields arise from (1,0)
string stretching from one D3-brane to another. Thus in the convention
described above, the electric charges carried by these gauge fields
take the form $\pm\wt\alpha$, $\pm\wt\beta$ and 
$\pm\wt\gamma=\pm(\wt\alpha+\wt\beta)$
with
\be \label{eform}
\wt\alpha = \pmatrix{1 \cr -1\cr 0}, \quad \wt\beta = 
\pmatrix{0\cr 1\cr -1}, \quad \wt\gamma = \pmatrix{1 \cr 0\cr -1}\, .
\ee
On the other hand in heterotic string theory on $T^4\times T^2$,
the SU(3)
gauge fields carry electric charge vectors $\pm\alpha$, $\pm\beta$
and $\pm\gamma$ given in \refb{eglim}. Thus we now have  a
correspondence between the charge vectors in the D3-brane
system to ones in heterotic string theory for states which are
charged only under the SU(3) subgroups in both theories.
In particular a state in the heterotic string theory carrying charges
$(p\alpha+q\beta, r\alpha+s\beta)$ will correspond to a 
three string junction carrying charges
\be \label{egen}
\wt Q = p \wt\alpha + q\wt\beta, \qquad 
\wt P = r \wt\alpha + s\wt\beta\, .
\ee

\begin{figure}
\leavevmode
\begin{center}
\epsfysize=4cm \epsfbox{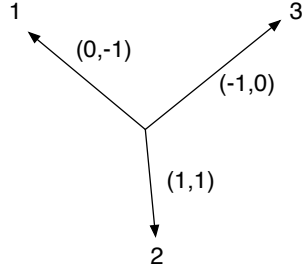}
\end{center}
\caption{The three string junction containing a $(0,-1)$ string,
a (1,1) string and a $(-1,0)$ string ending on D3 branes 1, 2 and 3
respectively.}
\label{f1}
\end{figure}

Let us now consider a three string junction in which a $(0,-1)$
string ends on the first D3-brane, a $(1,1)$ string ends on the second
D3-brane and a $(-1,0)$ string ends on the third D3-brane 
(see fig.~\ref{f1}). This
corresponds to the choice
\be \label{egen2}
\wt Q = \pmatrix{0 \cr 1 \cr -1} = \wt\beta,
\qquad \wt P = \pmatrix{-1 \cr 1\cr 0} = -\wt\alpha\, .
\ee
Thus in heterotic string theory on $T^4\times T^2$, the charge
vectors carried by this state will be $(\beta, -\alpha)=(Q_0,P_0)$
with $Q_0, P_0$ given in \refb{egen3}.

\begin{figure}
\leavevmode
\begin{center}
\epsfysize=4cm \epsfbox{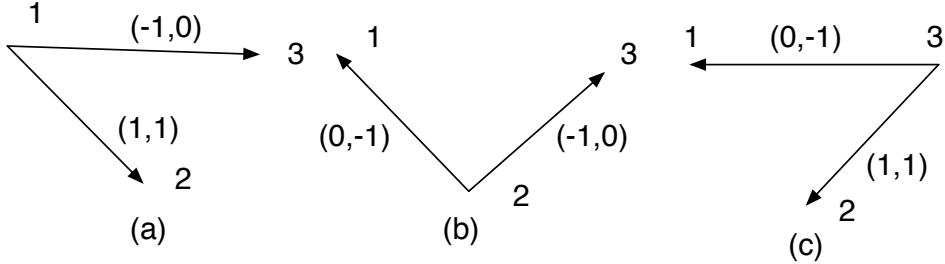}
\end{center}
\caption{Marginally stable three string junctions. In (a) the string
ending on the first D3 brane shrinks to zero size, in (b) the string
ending on the second D3-brane shrinks to zero size, and in (c) the 
string ending on the third D3-brane shrinks to zero size.}
\label{f2}
\end{figure}

We shall now  examine the domain in which
the three string junction exists and compare the result with the
one obtained from the dyon degeneracy formula in $\NN=4$
supersymmetric string theory.  The 3-string junction 
can become marginally unstable in one of three ways, corresponding
to shrinking one of the three strings to zero size\cite{9712211}
(see fig.\ref{f2}). 
Consider first the case
where the $(0,-1)$ string ending on the first D3-brane shrinks to
zero size. In this case the resulting configuration becomes
identical to that of a (1,1) string going from the first to the second
brane and a $(-1,0)$ string going from the first to the third brane.
According to our convention the former has charge vectors
\be \label{egen5}
\wt Q_1 = \pmatrix{-1 \cr 1 \cr 0} = \wt P, \quad 
\wt P_1 = \pmatrix{-1 \cr 1 \cr 0} = \wt P, 
\ee
whereas the latter has a charge vector
\be \label{egen6}
\wt Q_2 = \pmatrix{1\cr 0\cr -1} = \wt Q - \wt P, \qquad
\wt P_2 = \pmatrix{0\cr 0\cr 0} = 0\, .
\ee
Thus this particular wall of marginal stability corresponds to
the decay 
\be \label{egen7}
(\wt Q, \wt P) \to (\wt P, \wt P) + (\wt Q - \wt P, 0)\, .
\ee
Proceeding this way we see that the second wall of marginal stability,
corresponding to shrinking of the (1,1) string ending on the second
D3-brane to zero size, induces decay into an $(0,-1)$ string going
from the second to the first brane and a $(-1,0)$ string going
from the second to the third branes. This gives
\be \label{egen7a}
(\wt Q, \wt P) \to (0, \wt P) + (\wt Q, 0)\, .
\ee
Finally the third wall of marginal stability, corresponding to the
shrinking of the $(-1,0)$ string ending on the third brane to zero
size, induces decay into a  $(0,-1)$ string going from the third
to the first brane and a $(1,1)$ string going from the third to
the second brane. This gives
\be \label{egen7b}
(\wt Q, \wt P) \to (0, \wt P-\wt Q) + (\wt Q, \wt Q)\, .
\ee

Eqs.\refb{egen7}, \refb{egen7a} and \refb{egen7b} give the walls
of marginal stability bordering the domain in which the three string
junction under consideration exists. 
{}From \refb{edecay} we see that these are precisely the walls  which
border the domain $\RR$. Thus we see that the particular
three string junction under consideration
exists in the domain $\RR$, -- exactly as predicted by the 
dyon degeneracy
formula in string theory. The degeneracy 
formula in fact goes further and predicts
that these states will have degeneracy 1. This cannot be verified
directly using the three string junction picture since quantization of
such a configuration is difficult, but has been verified by working
in the gauge theory description of these 
states\cite{9907090,0005275,0609055}.

We can also consider a slightly different three string junction 
in which a $(0,1)$
string ends on the first D3-brane, a $(1,-1)$ string ends on the second
D3-brane and a $(-1,0)$ string ends on the third D3-brane. Following
the same procedure as in the previous case one finds that the
corresponding state in heterotic string theory on $T^4\times T^2$
has $Q^2=P^2=-2$, $Q\cdot P=1$. Thus this state exists
with degeneracy 1 in the domain $\LL$. An analysis identical to
the one described earlier shows that in the heterotic string
theory description the state ceases to exist as we cross
any of the walls of marginal stability bordering the domain $\LL$.
This leads to a definite prediction for the domain in the moduli
space of D3 brane configurations in which the three string
junction
exists. This can be verified explicitly in the same way as in the 
previous case.

Let us now consider the case of a
more general three string junction
configuration
where an $(m_1,n_1)$ string ends on the first D3-brane, an
$(m_2,n_2)$ string ends on the second D3-brane and an
$(m_3,n_3)$ brane ends on the third D3-brane. If 
$(m_1 n_3-m_3 n_1)=-1$ then this configuration may be obtained
from the one discussed earlier, -- containing a $(0, -1)$, $(1,1)$
and $(-1, 0)$ strings, -- by an S-duality transformation by the matrix
$\pmatrix{-m_3 & - m_1\cr -n_3 & - n_1}$. Thus the domain in
which it exists can be determined, -- 
both in the string junction 
description
and in the description as a dyon in the $\NN=4$ supersymmetric 
heterotic string
theory, -- by an S-duality transformation of the domain
in which the
configuration of $(0, -1)$, $(1,1)$
and $(-1, 0)$ string exists. Our earlier analysis for the latter
configuration now implies that the results in the two descriptions would
also agree for the more general configuration involving $(m_1, n_1)$,
$(m_2, n_2)$ and $(m_3, n_3)$ strings as long as 
$(m_1 n_3-m_3 n_1)=-1$. Similarly if $(m_1 n_3-m_3 n_1)=1$ then
we can relate this to the configuration of $(0,1)$, $(1, -1)$
and $(-1, 0)$ string via an S-duality transformation 
$\pmatrix{-m_3 & m_1\cr -n_3 & n_1}$, and the agreement between
the results based on dyon spectrum in string theory and three string
junction would follow as a consequence of a similar agreement for the
$(0,1)$, $(1, -1)$
and $(-1, 0)$ configuration.

What about the case when 
$(m_1 n_3 - m_3 n_1)\ne \pm 1$? 
It is
instructive to see what kind of charge vectors the general
$(m_1, n_1)$, $(m_2,n_2)=(-m_1 -m_3, -n_1-n_3)$,
$(m_3, n_3)$ string configurtion corresponds to in 
heterotic string theory.
In the string junction description the charge vectors are
\be \label{eth1}
\wt Q = \pmatrix{ m_1 \cr -m_1 - m_3 \cr m_3} = m_1 \wt\alpha
- m_3\wt\beta, \qquad
\wt P = \pmatrix{ n_1 \cr -n_1 - n_3 \cr n_3} = n_1 \wt\alpha
- n_3\wt\beta\, .
\ee
Thus in the heterotic string theory on $T^4\times T^2$
this would
correspond to the charge vectors
\be \label{eth2}
Q = m_1  \alpha
- m_3 \beta = \pmatrix{m_1 \cr -m_1-m_3 \cr -m_1 \cr m_3},
\qquad 
P = n_1  \alpha
- n_3 \beta = \pmatrix{n_1 \cr -n_1-n_3 \cr -n_1 \cr n_3}\, .
\ee
In this case it is easy to see that
\be \label{eth3}
\hbox{g.c.d.}(Q_i P_j - Q_j P_i; \quad i,j=1,2,3,4) = 
|m_1 n_3 - m_3 n_1|\, .
\ee
For $|m_1 n_3 - m_3 n_1|\ne 1$ these states are outside the duality
orbit of the state 
\refb{eseed}\cite{9804160,0702150,0708.1270,0401049}. 
Hence the currently known formula for
degeneracy of dyons in $\NN=4$ supersymmtric string theories 
do not have any information about the spectrum of these states.
Indeed, since one can now construct a non-premitive lattice
vector from an integer linear combination of $\vec Q$ and $\vec P$,
even the structure of the marginal stability walls change, --
in the decay $(Q,P)$ into $(ad Q-ab P, cd Q - cb P)$ and
$(-bc  Q + ab  P, -cd  Q+ad  P)$ the coefficient $a$, $b$, $c$, $d$
need not all be integers any more.

Different aspects of the relation between dyon spectrum 
in supersymmetric
gauge theories and string theories have been discussed in
\cite{appear}.

\noindent {\bf Acknowledgement}: I would like to thank Atish
Dabholkar and Suresh Nampuri for useful discussions and the
people of India for generous support to research in string theory.

\end{document}